\documentclass[epsf,twocolumn,preprintnumbers]{revtex4}
\usepackage{graphics}
\usepackage{graphicx}
\usepackage{dcolumn} 
\usepackage{bm}
\usepackage{color}
\usepackage{epsfig}
\newcommand{\ssoo}{Sr$_2$SrOsO$_6$}
\newcommand{\bnoo}{Ba$_2$NaOsO$_6$}
\newcommand{\bmro}{Ba$_2$MgReO$_6$}
\newcommand{\bzro}{Ba$_2$ZnReO$_6$}

\begin{document}
\title{Symmetry-protected Spinful Magnetic Weyl Nodal Loops and Multi-Weyl Nodes \\
in $5d^n$ Cubic Double Perovskites $(n=1,2)$
}
\author{Young-Joon Song$^1$}
\author{Kwan-Woo Lee$^{1,2}$}
\affiliation{
 $^1$Division of Display and Semiconductor Physics, Korea University, Sejong 30019, Korea\\
 $^2$Department of Applied Physics, Graduate School, Korea University, Sejong 30019, Korea
}
\date{\today}
\begin{abstract}
Using both an effective three-band model and {\it ab initio} calculations,
we have investigated various topological features in the cubic ferromagnetic $5d^{1,2}$ systems 
showing large spin-orbit coupling (SOC): Ba$_2$NaOsO$_6$, Sr$_2$SrOsO$_6$, and Ba$_2$$B$ReO$_6$ ($B$= Mg, Zn).
In the presence of time-reversal symmetry (${\cal T}$), 
spinless Dirac nodal loops linked to each other at the $W$ points appear in the mirror planes.
Remarkably, breaking ${\cal T}$
leads to spinful magnetic Weyl nodal loops (MWNLs) 
that are robust even at large SOC and correlation strength $U$ variation
due to the combination of mirror symmetry and broken ${\cal T}$.
Additionally, there are two types of magnetic Weyl points with chiral charges $|\chi|=1, 2$ 
along the $C_{4v}$ symmetry line,
and another type-II MWNL encircling the zone center, that are dependent on $U$.
Furthermore, 
the ferromagnetic Ba$_2$ZnReO$_6$ 
is an ideal  half semimetal with MWNLs and magnetic Weyl nodes at the Fermi level  
without the interference of topologically trivial bulk states. 
These systems give rise to a remarkably large anomalous Hall conductivity $\sigma_{xy}$
of up to 1160 ($\Omega$cm)$^{-1}$.
Our findings may apply widely for $t_{2g}$ systems 
with cubic (or slightly distorted) fcc-like structures.
\end{abstract}
\maketitle

\section{Introduction}
The rocksalt-like double perovskites host abundant exotic physical properties related 
to interactions among the spin, orbital, and lattice degrees of freedom.
They have been considered as potential candidates for various applications of 
spintronics, multiferroics, magnetocalorics, and even solar cells.
In heavy transition metal systems, the large spin-orbit coupling (SOC) 
leads to the unconventional relativistic Mott transition.\cite{bnoo,bnoo1}
When the systems contain magnetic ions, 
half-metallic ferromagnets (FMs) or antiferromagnets can emerge.\cite{wep97,LP08}
Recently, their topological characters have also begun to be investigated.\cite{pi17}

Recent progresses in topological physics are striking experimentally and theoretically. 
In addition to topological insulators, 
topological semimetals and metals have been intensively investigated, in particular in three-dimensional (3D) systems.\cite{ts1}
In 3D systems with both time-reversal (${\cal T}$) and inversion (${\cal P}$) symmetries,
linearly crossing bands form fourfold Dirac points (DPs).
Breaking either ${\cal P}$ or ${\cal T}$ (or both)\cite{rev-tsm-APX}  
splits a DP into a pair of Weyl points (WPs) with opposite chiralities, 
leading to a Fermi arc surface state connecting the pair of WPs.
Systems with these DPs or WPs show various abnormal transport properties like ultra-high mobility, 
extremely large magneto-resistance, and chiral anomaly.\cite{ts-exp1,ts-exp2}
Magnetic Weyl phases have been observed very recently in the FM Fe$_3$Sn$_2$\cite{magWeyl18} 
and Co$_3$Sn$_2$S$_2$\cite{magWeyl19-2,magWeyl19-3} with a kagome lattice, 
and the FM CeAlGe with a nonsymmorphic tetragonal structure\cite{magWeyl19},
whereas a ${\cal P}$-broken Weyl phase was observed just a few months after its prediction.\cite{WSnoP}

In addition to the ordinary WP with a chiral charge $|\chi|=1$, corresponding to the number of the Fermi arcs,
multi-WPs with higher chiral charges $\chi$ have also been proposed,\cite{c.fang12,brad16}
and some of them have been experimentally observed.\cite{DW-exp19}
Weyl fermionic states with $|\chi|=2$ can emerge in quadratically touching points,\cite{huang16,z.zhu18}
in linearly crossing four-bands,\cite{tang17} and in linearly crossing two-bands with a flat band crossing the nodal point\cite{brad16}.
The first two are called double WPs, and the last one is the spin-1 chiral fermion.
In the absence of ${\cal T}$,
a single-spin double WP can appear along the $C_{4,6}$-rotation axis,
when two bands touch quadratically in the direction perpendicular to the rotational axis.\cite{jpn02,QAHE11,c.fang12}
For 3D crystals with screw symmetry $n_m (n=4,6)$, multi-Weyl nodes of $|\chi|=2,3$ when ${\cal P}*{\cal T}$ is broken 
have also been suggested\cite{screw17,screw19}.
Systems with multi-WPs were suggested to show distinctive optical and electromagnetic responses.\cite{min17,huang17}

The Dirac (or Weyl) nodes can form a one-dimensional nodal line or loop,\cite{dnl1,dnl2} 
a two-dimensional(2D) nodal surface,\cite{nodalsf1,nodalsf2,nodalsf3,nodalsf4}
or various kinds of 3D nodal links.\cite{nodalchain1,nodalchain2,nodalchain3,nodalchain4}
The nodal lines or loops (NLs) lead to a less dispersive surface band connecting two nodes, dubbed the drumhead state,\cite{dnl1}
giving rise to a large surface density of states.
One may expect instabilities toward surface superconductivity or magnetism,\cite{drum,lado19}
when NLs are close to the Fermi energy $E_F$.
Additionally, similar to the type-II WP,\cite{type2WS}
a Weyl line with a tilted Weyl cone on the NLs has been proposed.\cite{type2nl16,type2nl17} 
In contrast to the conventional Weyl NL,
distinctions in the magnetic and optical responses in the type-II NL like a collapse of the Landau levels  
resulting from the effective masses of the two bands having the same signs in the tangential direction
have been suggested.\cite{type2nl17,type2nl18}  
Recently, Chang {\it et al.} observed the type-II NL in Mg$_3$Bi$_2$ with a La$_2$O$_3$-type structure
through angle-resolved photoemission spectroscopy measurements.\cite{type2nl19}

As is often the case in topological properties\cite{rev-tsm-APX},
the effects of SOC are one of the vital ingredients in these NLs, 
regardless of their types.
Systems with NLs robust against SOC are very rare, since SOC often opens a gap in these NLs.
Additional crystalline symmetries can protect these NLs.\cite{dnl1,ts-rot1,ts-rot2,ts-m1,ts-m2,ts-ns1,ts-ns2} 
Alternatively, in the broken-${\cal T}$ case, NLs protected by a mirror symmetry are robust against SOC.\cite{c.fang16,y.kim18}
However, ${\cal T}$-broken NLs have been proposed in relatively limited systems\cite{magloop1,magloop2,magloop3},
and no Weyl NL without ${\cal T}$ has been realized yet. 

In this paper, we will address various topological features of the cubic double perovskite osmates \bnoo~and \ssoo,
and rhenates Ba$_2$$B$ReO$_6$ ($B$=Mg, Zn), using both model and {\it ab initio} approaches.
The osmates are rare examples of (nearly) cubic FM insulators 
with Curie temperatures $T_C\approx$ 8 K (\bnoo)\cite{bnoo-exp1,bnoo-exp2} and 1000 K (\ssoo)\cite{ssoo-exp}.
The FM rhenates have $T_C\approx10-20$ K.\cite{bzro}
These systems show large $t_{2g}-e_{g}$ crystal-field splittings of several eV 
and large $p-d$ hybridization gaps of roughly a few eV, 
leading to an isolated and partially filled $t_{2g}$ manifold, as observed in \bnoo\cite{bnoo}.
Thus, these systems can be described well by an effective three-band model of the $t_{2g}$ manifold.
Remarkably, in these systems, the interplay of strong SOC on the order of 0.3 eV, moderate correlation strength of a few eV,
and crystalline symmetries significantly affects their topological characters, as will be discussed below.
Our results show various rare and interesting topological properties: 
spinful magnetic Weyl nodal loops (MWNLs) of two different types, multi-Weyl nodes of chiral charge $|\chi|$=2,
half semimetals with MWNLs and WPs at $E_F$, and high anomalous Hall conductivity $\sigma_{xy}$. 
A system that possesses both half-metallic and topological characters near $E_F$
is a promising candidate for spintronics applications with small energy dissipation and high-speed performance.\cite{nexus19,r.zhang20}

The rest of this paper is organized as follows.
The computational approaches are described in Sec. II.
In Sec. III, the results obtained from the effective three-band model are presented.
The topological characters of the osmates are discussed in Sec. IV.
The ideal magnetic Weyl half-metallic character in the rhenates is presented in Sec. V.
Then, a brief summary follows in Sec. VI.

\begin{figure}[tbp]
{\resizebox{8cm}{3.5cm}{\includegraphics{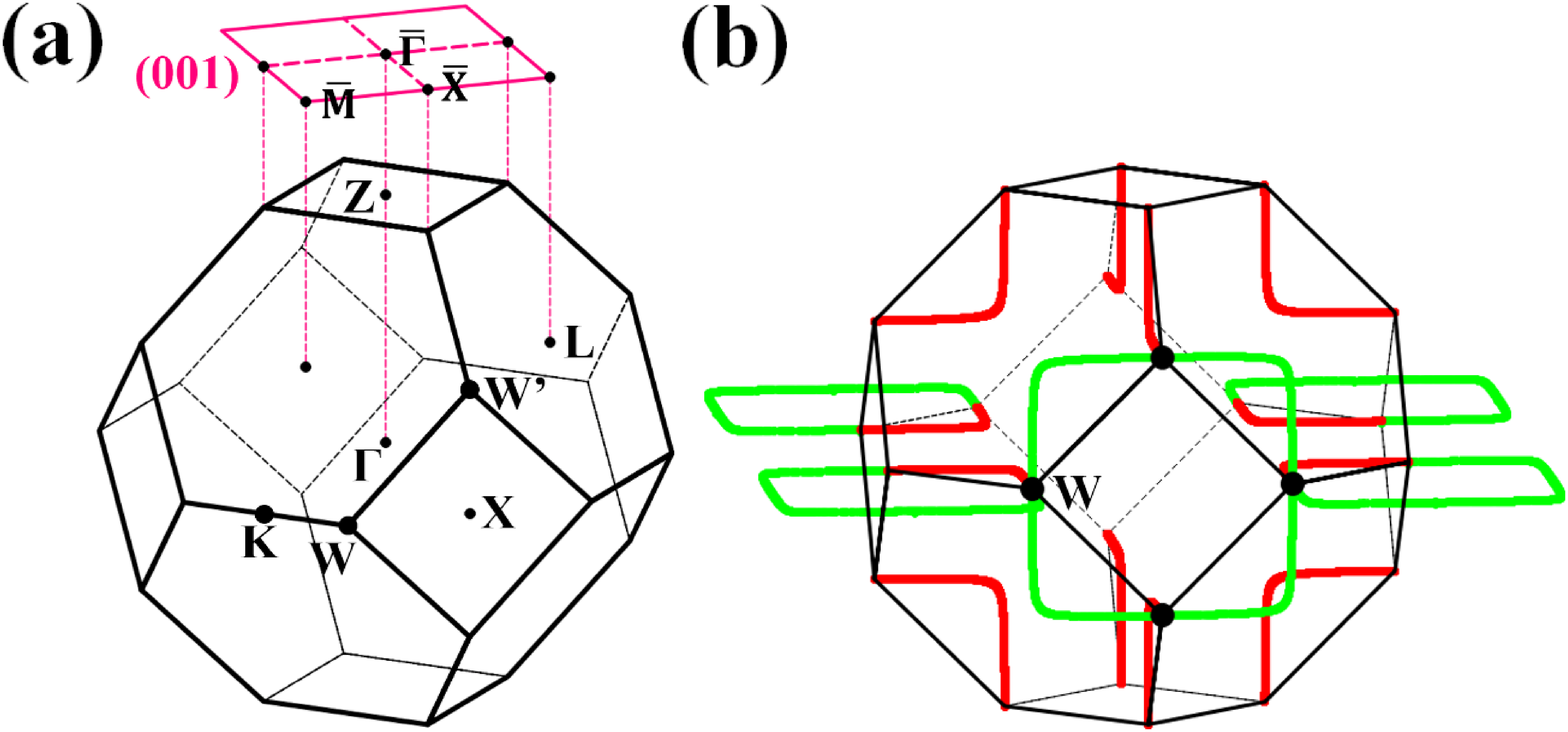}}}
\caption{
(a) Brillouin zones (BZs) of bulk and (001) surfaces with high-symmetry points
of the fcc structure.
(b) Spinless nodal loops, penetrating the BZ, in the ${\cal T}$-preserved case.
The light (green) colored loops are represented in the extended zone scheme for better visualization.
As shown on the front face, each loop is linked at the $W$ points
at the middle of each side.
}
\label{bz}
\end{figure}

\section{Calculation Methods}
In the cubic double perovskites (space group: $Fm\bar{3}m$),
the experimental lattice parameters of 8.287 \AA~for \bnoo\cite{bnoo-exp1,bnoo-exp2},
8.23 \AA~for \ssoo\cite{ssoo-exp}, 8.07 \AA~ for \bmro\cite{bzro}, and 8.10 \AA~ for \bzro\cite{bzro} were used. 
The internal parameters of the oxygen positions were relaxed 
using the generalized gradient approximation (GGA)\cite{gga} until the force was smaller than 1 mRy. 
We obtained values consistent with the experiments\cite{ssoo-exp,bnoo-exp2,bzro}.

We performed {\it ab initio} calculations based on the GGA exchange correlation functional
implemented in the accurate full-potential all-electron code {\sc wien2k}.\cite{wien2k}
The effects of correlation and SOC in these $5d$ systems 
were treated with the GGA+SOC+$U$ approach.
In the GGA+SOC calculations, we chose the magnetization direction to be along the (001) axis, 
which is the highest symmetry direction, because of the tiny observed magnetic anisotropy.\cite{ssoo-exp}.
To obtain an insulating state,
we used the effective $U_{eff}=U-J=3-6$ eV, which is somewhat dependent on the specific systems.
Here, $U$ and $J$ are the on-site Coulomb repulsion and Hund's integral, respectively.
Such a large critical value of $U^c_{eff}$ for the metal-insulator transition 
is necessary due to the strong $p-d$ hybridization in these systems,
as discussed by Lee and Pickett,\cite{bnoo}
although a much smaller $U^c_{eff}$ is usually expected for such an extended $5d$ orbital.
We also carried out calculations with a separate $U$ and $J$ scheme, and obtained similar results.
We will hence present only the results obtained from the $U_{eff}$ scheme.

In {\sc wien2k}, the fcc-like Brillouin zone (BZ) shown in Fig. \ref{bz}(a) 
was sampled by a dense regular $k$ mesh containing 1992 points in the irreducible wedge.
The basis size was determined by $R_{mt}K_{max}=7$ and the augmented-plane-wave sphere radii (in atomic units):
Os 2.1, O 1.4, Ba 2.3, Na 2.2 for \bnoo; Os 2.1, O 1.45, Sr 2.1 for \ssoo;
Re 2.2, O 1.43, Ba 2.3, Zn 2.15 for \bzro; Re 2.2, O 1.43, Ba 2.3, Mg 1.7 for \bmro.

In addition to the effective three-band model that will be discussed in the next section,
we obtained the Hamiltonian for the $t_{2g}$ states near $E_F$ 
using the maximally localized Wannier functions technique, implemented 
in {\sc wannier90}\cite{w90}, to investigate the topological characters.
To prepare the input files for {\sc wannier90}, the {\sc wien2wannier} program was used.\cite{w2w}
The full BZ was sliced by a 17$\times$17$\times$17 $k$ mesh to obtain the Wannierized Hamiltonian 
with six initial orbitals.
With these Hamiltonians, the surface states were obtained using the iterative Green's-function method\cite{surfgr},
implemented in the {\sc wanniertools} package\cite{wt}.

The anomalous Hall conductivity (AHC) $\sigma_{xy}$ was calculated with a fine $300\times300\times300$ $k$ mesh.
The AHC $\sigma_{xy}$ is directly related to the Berry curvature via the Kubo formula\cite{ahc-eq1,ahc-eq2}
\begin{eqnarray}
\sigma_{xy} = -\frac{e^2}{\hbar}\sum_{n}\int\frac{d^3k}{(2\pi)^3}\Omega_{xy}^{ n}(\mathbf{k})f_n,
\end{eqnarray}
where $f_n$ is the Fermi-Dirac distribution function.
Measuring AHC is one of the recent experimental approaches to efficiently observe the topological characters.\cite{magWeyl18,liu2018}
The Berry curvature $\Omega_{xy}^{ n}$ of the $n$th band is given by
\begin{eqnarray}
\Omega_{xy}^{ n}(\mathbf{k}) = -2{\rm{Im}}\sum_{m\neq n}\frac{<\psi_n({\mathbf{k}})|
\frac{\partial \mathcal{H}}{\partial k_x}|\psi_m({\mathbf{k}})> 
<x\rightarrow y>^\ast
}
{\left[\varepsilon_m(\mathbf{k}) - \varepsilon_n(\mathbf{k})\right]^2}.
\end{eqnarray}
Here, $\psi_m({\mathbf{k}})$ and $\varepsilon_n(\mathbf{k})$ 
are the $n$th Bloch eigenstates and eigenvalues of the Hamiltonian ${\mathcal{H}}(\mathbf{k})$.

\section{Results of the three-band model}
First, we describe the results obtained from the effective three-band model calculations.
The model ${\cal H}_{TB}$ for the isolated $t_{2g}$ manifolds can be well described by  
the three nearest-neighbor (NN) hopping parameters of $t_\sigma$, $t_\pi$, and $t_\delta$, 
as previously shown in \bnoo\cite{bnoo}.
We also include SOC with a strength of $\xi_{SOC}$ and the exchange splitting $\Delta_{ex}$ term.
The latter term is included to investigate the ${\cal T}$-broken case.
The full Hamiltonian ${\cal H}_{t_{2g}}$ is hence given by
 \begin{eqnarray}
{\cal H}_{t_{2g}} = {\cal H}_{TB} + \xi_{SOC}\vec{L} \cdot \vec{S} + \Delta_{ex}\sigma_z,
\label{tb}
\end{eqnarray}
where $\sigma_z$ is the Pauli matrix.
The effects of correlation 
will be discussed in the realistic systems (see below).


\begin{figure}[tbp]
{\resizebox{8cm}{8cm}{\includegraphics{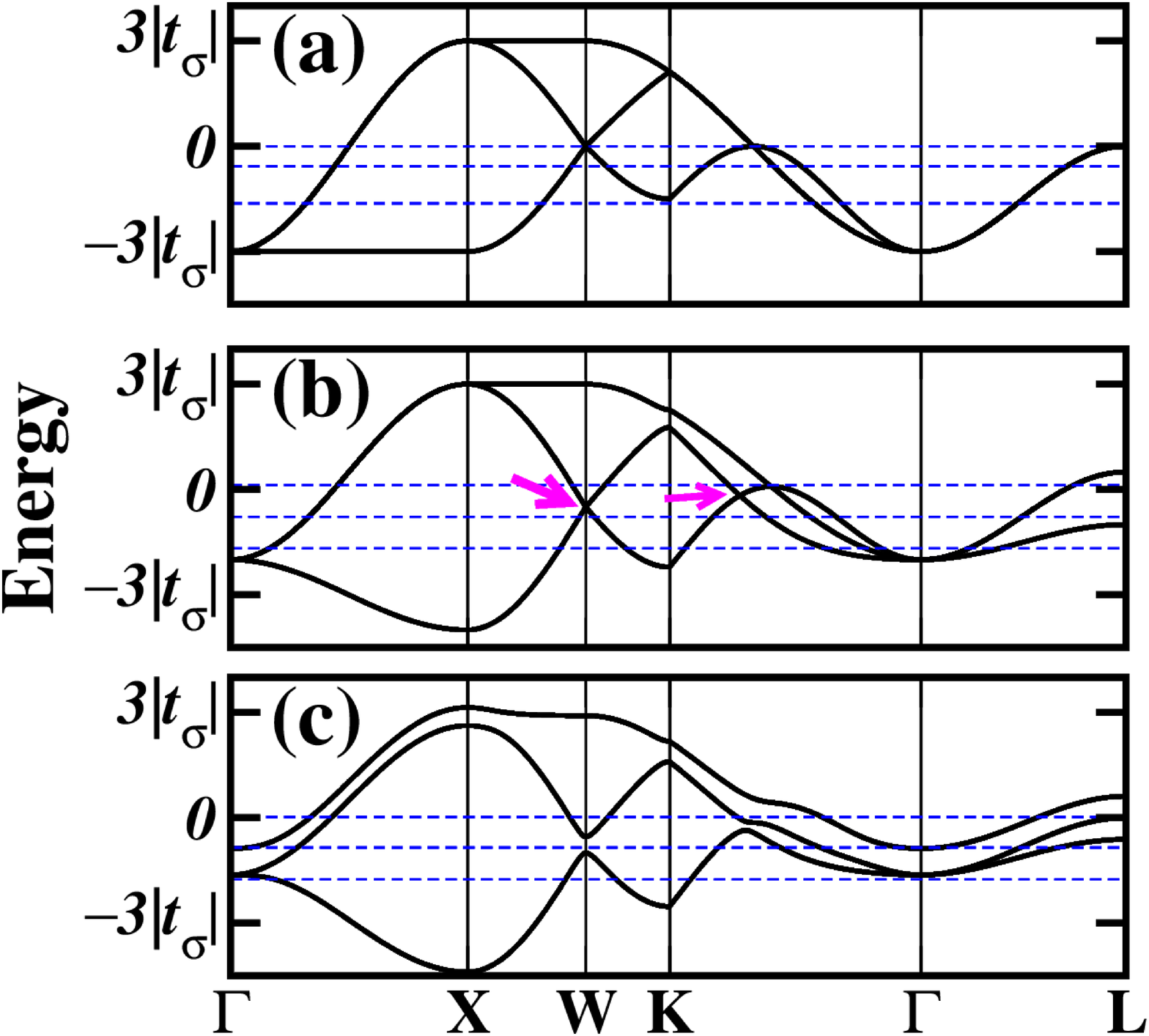}}}
\caption{
Evaluation of ${\cal T}$-preserved band structures of the three-band model,
when (a) only $t_\sigma$, (b) both $t_\sigma$ and $t_\pi$, and 
(c) additional $t_\delta$ and SOC of $\xi_{SOC}=0.2$ eV were included. 
In these plots, $t_\sigma=-0.2$ eV is used for illustration. 
In (b) and (c) we also used $t_\pi = -t_\sigma/4$ and $t_\delta=t_\pi/2$.
The arrows in (b) denote DPs.
The horizontal (blue) dashed lines indicate each Fermi level $E_F$ of the $d^n$ configuration ($n=1, 2, 3$).
}
\label{tb-band1}
\end{figure}

\subsection{Presence of time-reversal symmetry}
For the ${\cal T}$-preserved case, i.e., neglecting $\Delta_{ex}$ in the model above,
we first consider the effects of the NN hopping parameters, and then of SOC.

As the hopping parameters are varied, the band structures 
are given in Figs.  \ref{tb-band1}(a) and \ref{tb-band1}(b).
The inclusion of only $t_\sigma$ yields a linearly crossing band at the $W$ point
and a twofold (neglecting spin) degenerate band along the $K-\Gamma$ line.
Adding $t_\pi$ makes the flat band along the $\Gamma-X$ line dispersive, 
as shown in Fig. \ref{tb-band1}(b).
Remarkably, the twofold degeneracy along the $K-\Gamma$ line is removed,
resulting in another linearly crossing band in the middle of the $K-\Gamma$ line.
The crossing bands marked by arrows in Fig. \ref{tb-band1}(b) 
lead to spinless Dirac NLs penetrating the BZ on the mirror planes of $M_x$, $M_y$, and $M_z$.
As shown in Fig. \ref{bz}(b), these loops with rounded square shapes in the fcc BZ
are of the outer nodal chain type\cite{nodalchain1} that are linked at the $W$ points, i.e., 
at the halfway point of each side.
Note that these loops are protected by the mirror and $C_4$ symmetries, 
as discussed by a two-band model.\cite{twobands} 
As expected, the effect of small $t_\delta$ on the topological characters is negligible.

Once the SOC is included, as observed previously\cite{dnl1}, 
a gap opens at every node [see Fig. \ref{tb-band1}(c)].
However, this gap opening leads to another topological character, 
as will be discussed in the real systems (see below).
Note that the gap lies near the $d^2$ $E_F$ in the ${\cal T}$-preserved case.


\begin{figure}[tbp]
{\resizebox{8cm}{6cm}{\includegraphics{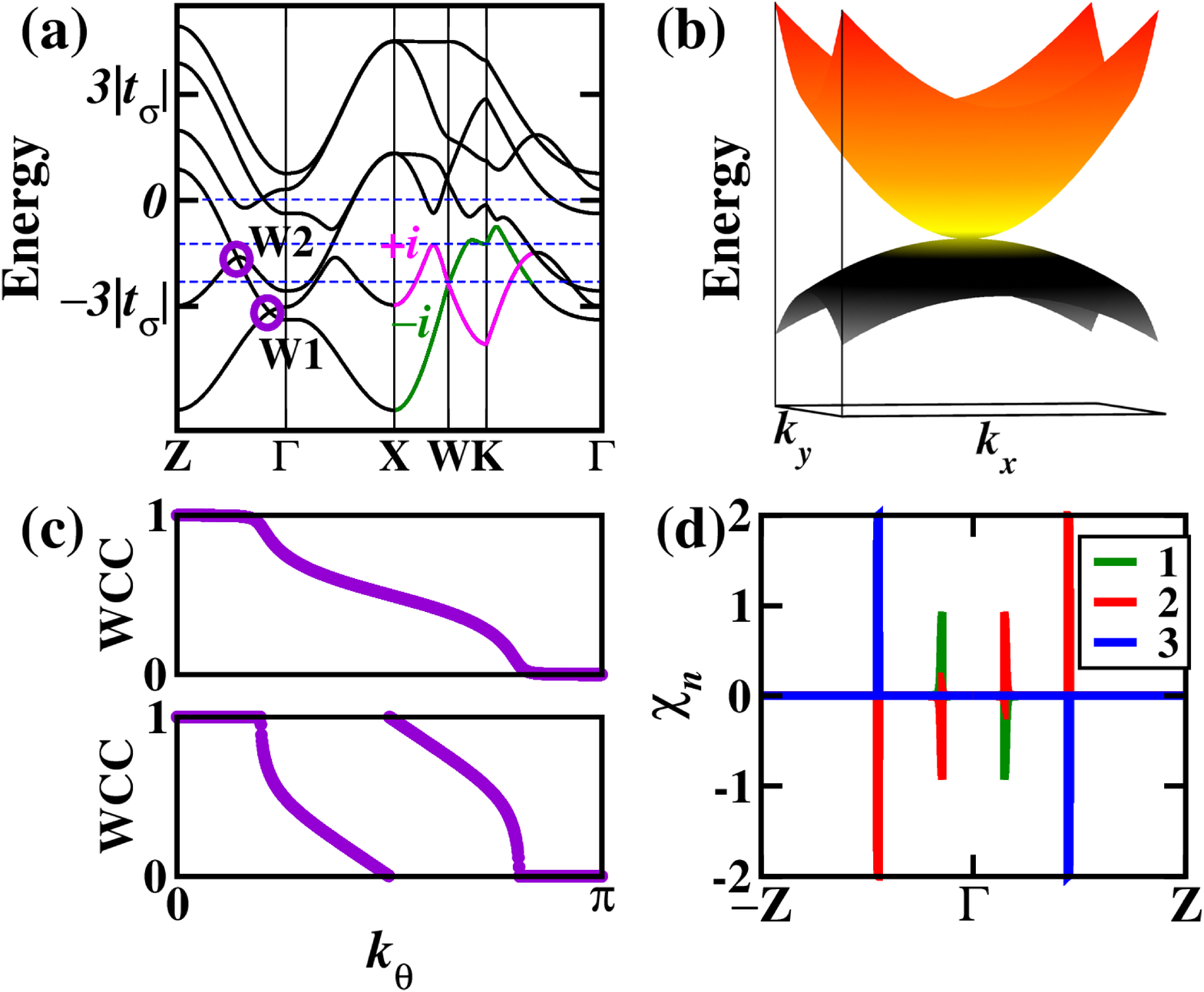}}}
\caption{
(a) Band structure of the three-band model, when breaking ${\cal T}$ by $\Delta_{ex}=0.3$ eV
in Fig. \ref{tb-band1}(c). 
Along the $\Gamma$-$Z$ line, two types of WPs, denoted by W1 and W2, emerge. 
The $\pm i$  indicate the mirror eigenvalues of the two bands that produce nodal loops.
The Fermi energies of $d^n$ ($n=1, 2, 3$) are denoted by the (blue) dashed lines.
(b) Band dispersions around the W2 point on a plane perpendicular to the $k_z$ axis, 
implying an exotic Weyl node.
(c) Plots of the Wannier charge center (WCC) of the (top) W1 and (bottom) W2 points. 
(d) Band-resolved chiral charge $\chi_n(k_z)$ along the $k_z$ axis for the lowest three bands 
(labeled in order of lower energy).
}
\label{tb-band2}
\end{figure}

\subsection{Broken time-reversal symmetry}
The ${\cal T}$-broken case is treated with the exchange splitting $\Delta_{ex}$, 
which converts each DP into a pair of WPs separated by $\Delta_{ex}$ (not shown here).
As a result, each Dirac NL splits into a pair of MWNLs on each mirror plane, 
again separated by $\Delta_{ex}$.

Next, in addition to $\Delta_{ex}$, we apply SOC with a magnetization direction of (001).
Hence, only the mirror symmetry $M_z$, which is perpendicular to the $k_z$ axis, 
remains, whereas $M_x$ and $M_y$ are vanished.
As shown in the band structure in Fig. \ref{tb-band2}(a),
the SOC leads to the other remarkable topological features of spinful MWNLs and two types of WPs.
The two bands linearly crossing at the $W$ point and in the middle of the $K-\Gamma$ line 
have opposite mirror eigen values of $\pm i$ about $M_z$,
as depicted in Fig. \ref{tb-band2}(a). Thus, they cannot be hybridized with each other by SOC(001)
on the $M_z$ (i.e., $k_z=0$) plane, leading to a gapless state.
As a result,
in contrast to the ${\cal T}$-preserved case,
the MWNLs survive on the $M_z$ plane, even after SOC(001) is applied.
The combination of mirror and broken ${\cal T}$ symmetries
protect these spinful NLs\cite{c.fang16,rev-tsm-APX,y.kim18}, 
which have been very rarely predicted in realistic compounds.
 
Another interesting feature appears along the the $\Gamma-Z$ line with the $C_{4v}$ symmetry.
Below the $E_F$ of the $d^3$ configuration, 
there are two linear band crossings denoted by W1 and W2 in Fig. \ref{tb-band2}(a),
which suggest magnetic WPs.
The positions of the W1 and W2 points at (0,0,$k_{zc}$)$\frac{\pi}{a}$ are determined by hopping parameters, 
$\Delta_{ex}$ and $\xi_{SOC}$, as given below:
\begin{eqnarray}
 k_{zc}(W1)&=& \cos^{-1}[1+\frac{\Delta_{ex}\xi_{SOC}}{(2\Delta_{ex}+\xi_{SOC})t}], \\
 k_{zc}(W2)&=& \cos^{-1}[1+\frac{2\Delta_{ex}}{t}],
\label{wp}
\end{eqnarray}
where $t=3t_\sigma-2t_\pi-t_\delta$ is negative for the $t_\sigma$.
Here, the $k_{zc}$ values are 0.1461 and 0.4448 for W1 and W2, respectively.
For $\Delta_{ex}=0$ these points merge into one point at the zone center, 
as obtained in the ${\cal T}$-preserving case. 
Note that $k_{zc}(W2)$ is independent of the SOC strength.

In particular, as shown in Fig. \ref{tb-band2}(b), 
the band touching with opposite effective masses at W2 is unusually quadratic
in the direction perpendicular to the rotation axis,
implying an exotic Weyl node.\cite{jpn02,QAHE11,c.fang12}
To uncover the origin of these nodal points W1 and W2,
their Wannier  charge  centers (WCCs),\cite{wcc14} implemented in the {\sc wanniertools} package\cite{wt},
were calculated.
Figure \ref{tb-band2}(c) shows the results obtained for W1 and W2.
As expected from the unusual dispersion, 
the WCC of W2 indicates a multi-WP with a chiral charge $|\chi|=2$,
whereas that of W1 indicates a conventional WP of $|\chi|=1$.

In order to confirm these chiral charges, 
we consider a minimal 2D Hamiltonian 
obtained by expanding the Hamiltonian ${\cal H}_{TB}$ of Eq. (\ref{tb}) to second order.
Ignoring the spin, the Hamiltonian $h(q_x,q_y$) is given by 
\[ \left(\begin{array}{ccc}
a - Aq_{x}^2 - Bq_{y}^2  & 0   & Dq_{x}   \\ 
0   & a - Bq_{x}^2 - Aq_{y}^2   & Dq_{y}   \\
Dq_{x}   & Dq_{y} & b - C(q_{x}^2 + q_{y}^2)
\end{array}\right) \]
in the basis of $\{d_{yz},d_{xz},d_{xy}\}$.
Here, the coefficients of the quadratic terms are 
$A=t_1(1+c_{kc})$, 
$B=t_1 + t_2c_{kc}$,
and $C=t_2 + t_1c_{kc}$,
whereas that of the linear term is $D=t_3s_{kc}$.
The zeroth-order terms are 
$a=\varepsilon_{0} + t_1(1 + c_{kc}) + t_2c_{kc}$
and $b=\varepsilon_{0} + t_2 + 2t_1c_{kc}$,
where $c_{kc}=\cos(k_{zc}\frac{\pi}{a})$ and $s_{kc}=\sin(k_{zc}\frac{\pi}{a})$
The hopping parameters are $t_1=2(t_\pi+t_\delta)$, $t_2=3t_\sigma+t_\delta$, 
and $t_3=-2(t_\pi-t_\delta)$. 
Using Eq. (\ref{tb}) with this Hamiltonian, we calculated the band resolved chiral charge $\chi_n(k_z)$
directly from the summation of the Berry curvatures $\Omega_z(\mathbf{k})$ 
by using the gauge-invariant representation of the Chern number 
on each plane of a sliced BZ along the $k_z$ direction.\cite{chern05}
The chiral charge $\chi$ at a WP is the summation of $\chi_n$ over all the bands below the WP.
The obtained results for the three bands forming W1 and W2 along the $k_z$ axis
are given in Fig. \ref{tb-band2}(d),
indicating one pair of ordinary WPs and another pair of multi-WPs with $|\chi|=2$.
Each pair has opposite chiralities.
Therefore, this magnetic phase hosts symmetry-protected MWNLs robust under SOC,
and two types of magnetic WPs along the $C_{4v}$ rotational axis.

\section{Application to Osmates}

\begin{table}[b]
\caption{On-site energy $\varepsilon_0$ and hopping parameters 
of \bnoo, \ssoo, \bzro, and \bmro, in units of meV.
The values for \bnoo~ are identical to those of Ref. \cite{bnoo}.
The $t_{\pi}^\prime$ indicates the NNN $dd\pi$ parameter.
}
\begin{center}
\begin{tabular}{ccccccc}\hline\hline
 ~~~&~~~$\varepsilon_0$~~& ~~~~$~t_{\sigma}~$~~~&~~~$t_{\pi}$~~~&~~~$t_{\delta}$~~~& ~~~$t_{\pi}^\prime$ \\\hline
\bnoo   & 202 & -121 & 64   & 24  &  0.0  \\
\ssoo   & 91  & -197 & 35.5 & 4.5 & -11.5 \\
\bzro   & 188  & -130 & 40 & 14 & 13.7 \\
\bmro   & 229  & -158 & 37.4 & 14 & 3 \\
\hline \hline
\end{tabular}
\end{center}
\label{t1}
\end{table}

\begin{figure*}[!tbh]
{\resizebox{16cm}{6cm}{\includegraphics{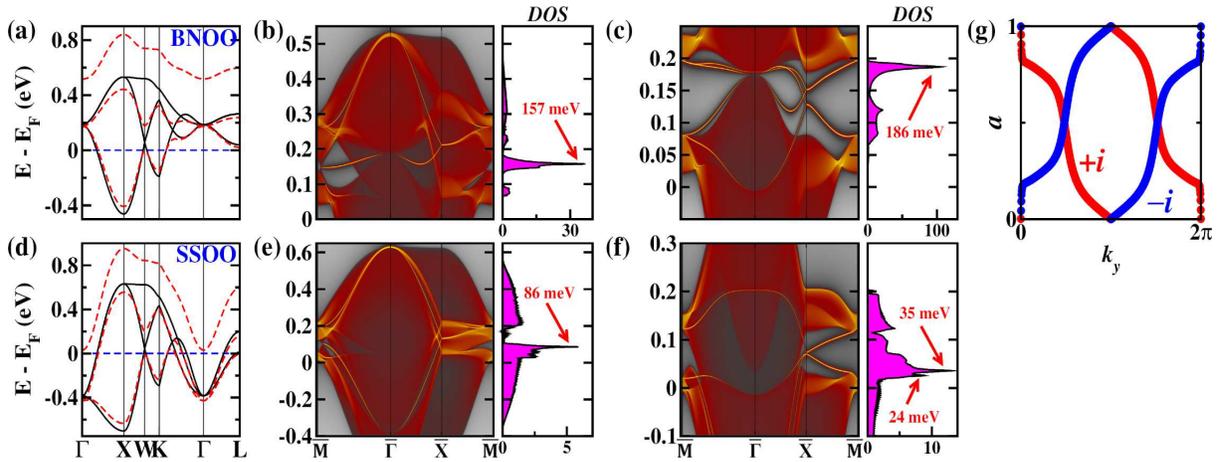}}}
\caption{
Band structures and surface spectra with densities of states (DOSs) in nonmagnetic \bnoo~(top) and \ssoo~(bottom).
These contain only the isolated and partially filled $t_{2g}$ manifolds.
(a),(d) Both GGA (black-solid)  and GGA+SOC (red-dashed) band structures.  
The band structure of \bnoo~ is consistent with the previous one in Ref. \cite{bnoo}.
Surface spectra and DOSs are shown without SOC in (b) and (e), and with SOC in (c) and (f).
The surface DOSs are given in arbitrary units.
The surface states lead to sharp peaks in the surface DOSs.
In the surface spectra plots, obtained in the conventional cubic cell,
the strong yellowish lines represent the surface states.
(g) WCCs of \bnoo~ at the $k_z=0$ plane,
 indicating the mirror Chern number ${\cal C}_M=2$.
}
\label{nmband}
\end{figure*}

The results obtained from the effective model suggest 
that cubic fcc-type systems with abundant crystalline symmetries
are promising candidates for diverse interesting topological features depending on ${\cal T}$ and SOC. 
As realistic examples, we investigated four cubic double-perovskite $5d^n$ FMs
possessing large SOC: $5d^1$ \bmro, \bzro, and \bnoo~and $5d^2$ \ssoo.
At the nonmagnetic (NM) GGA level, the isolated $t_{2g}$ manifolds of these systems have 
very similar structures, except for $E_F$ and bandwidth (see below).
These can be well described by the effective three-band model of Eq. (\ref{tb})
with three NN hopping parameters and one next-nearest-neighbor (NNN) 
hopping parameter that are given in Table \ref{t1}.
In this section, we will focus on the osmates, and the rhenates will be discussed in the next section.

\subsection{Nonmagnetic phases}
Figures \ref{nmband}(a) and \ref{nmband}(d) show the blowup NM band structures, including only the isolated $t_{2g}$ orbital,
of \bnoo~and \ssoo~in the GGA level without and with SOC.
Both band structures are very similar to each other, but the bandwidth of \ssoo~is about 
twice that of \bnoo~despite their similar volumes. 
As shown in Table \ref{t1}, 
the difference is mainly reflected in $t_\sigma$, which is about $\frac{2}{3}$ larger in \ssoo.
Note that an additional small NNN hopping parameter $t^\prime_\pi\approx-11$ meV
is necessary for describing the band structure of \ssoo~well.
$\xi_{SOC}\approx 0.3$ eV fits the SOC cases of both systems well.

To figure out the topological characters of these osmates, 
we analyzed the band structures in detail.
In the absence of SOC, both systems show 
a few fourfold degenerate points (including spins): 
two linearly crossing bands (DPs) at the $W$ point and along the $K-\Gamma$ line, 
and a quadratic touching point at the $\Gamma$ point. 
Notably, in spite of the different fillings,
the DPs at $W$ lie several tens of meV directly above $E_F$ in both systems.
Consistent with the results of our model calculations, 
the DPs generate Dirac NLs on the three mirror planes, as shown in Fig. \ref{bz}(b).
In the surface spectral functions, shown in Figs. \ref{nmband}(b) and \ref{nmband}(e), 
these spinless Dirac NLs lead to a (less dispersive) drumhead state 
visible along the $\bar{M}-\bar{\Gamma}$ line at about 150 meV (\bnoo) and 80 meV (\ssoo).
In \bnoo~the state emerges without bulk bands,
whereas that of \ssoo~merges into the bulk bands. 
This distinction is due to the fact that 
the energy $E_\Gamma$ is higher than $E_W$ only for \bnoo. 
The energy difference between the nodes at the $\Gamma$ and $W$ points is given by
$\Delta E=E_W-E_\Gamma\approx 3(|t_\sigma|-2t_\pi)$, neglecting the small parameters.

The inclusion of SOC leads to a large gap of 0.2 eV at the DPs,
resulting in the vanishing of the fourfold degeneracies in the whole BZ 
except for the quadratically touching bands at the $\Gamma$ point that are protected by the $O_h$ symmetry 
[see Figs. \ref{nmband}(a) and \ref{nmband}(d)].
The band structures on the $k_z=\pi/a$ planes in both systems can thus be considered as insulating subspaces.\cite{pbpd3}
The corresponding surface spectra of \bnoo~and \ssoo~are displayed in Figs. \ref{nmband}(c) and \ref{nmband}(f), respectively.
In \bnoo, most surface states, including the Dirac cones around the $\bar{X}$ point, appear 
in the range of 0.07 to 0.2 eV without any bulk bands.
However, most surface states of \ssoo~are accompanied by bulk bands, except on the $\bar{X}-\bar{M}$ line,
since $E_\Gamma$ is lower than $E_W$ by 0.4 eV. 
To uncover the origin of the surface states,
we calculated the $Z_2$ number on the insulating subspaces using the WCC approach,\cite{wcc14}
but the obtained $Z_2=0$ implies a non-ordinary $Z_2$ topological insulating phase.
Additionally, the mirror Chern number ${\cal C}_M=({\cal C}_{+i}-{\cal C}_{-i})/2=2$ was calculated with the WCC approach,
where ${\cal C}_{\pm i}$ denote the individual Chern numbers of the eigenstates with eigenvalues $\pm i$.\cite{teo08} 
As shown in Fig. \ref{nmband}(g), this indicates that $|C_M|=2$, consistent with the two pairs of surface states 
that are clearly visible along the $\bar{\Gamma}-\bar{X}$ line in Fig. \ref{nmband}(c).
Thus, the insulating subspace can be considered as a topological crystalline insulating phase
protected by the $C_{4v}$ symmetry.\cite{tci,ttn}
It is worth noting that a less dispersive surface state in \ssoo~lies very close to $E_F$ 
along the $\bar{M}-\bar{\Gamma}$ line, so leading to a large DOS at about 20 meV. 
Thus, an instability due to the large DOS may account for the experimentally observed high $T_C\sim$1000 K in \ssoo.

\begin{figure}[tb]
{\resizebox{8cm}{4cm}{\includegraphics{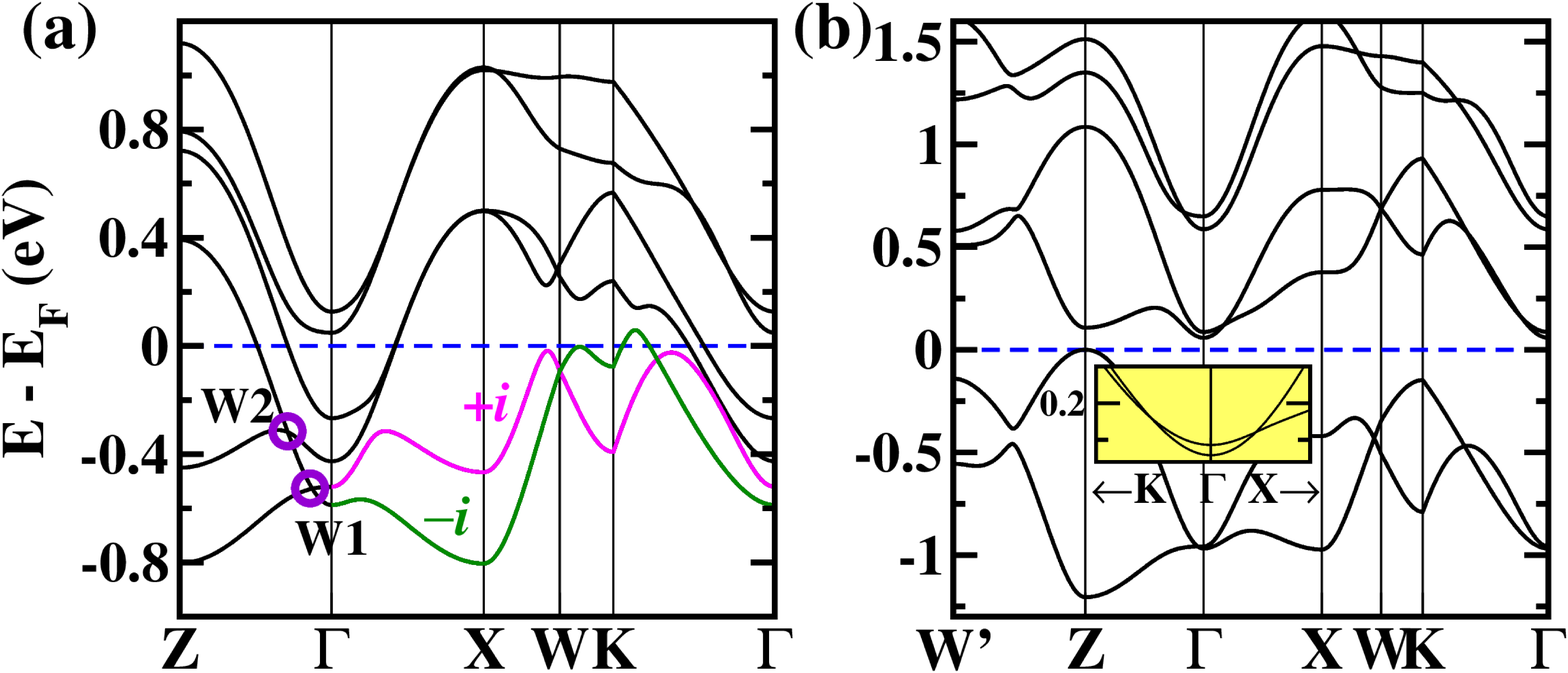}}}
\caption{
FM band structures containing only the $t_{2g}$ manifold of \ssoo~in GGA+SOC(001)+$U$ 
with (a) $U_{eff}=0$ and (b) 4 eV.
The latter represents the observed insulating phase.
The former is very similar to the results of our model calculations given in Fig. \ref{tb-band2}(a).
In (a), W1 and W2 indicate WPs with the chiral charges of $|\chi|=1, 2$, respectively, as in Fig. \ref{tb-band2}(a).
Inclusion of correlation leads to another nodal loop around the $\Gamma$ point, shown in the inset of (b).
In (b), a pair of WPs appears only at 0.7 eV along the $\Gamma-Z$ line.
Here, all symmetry points, except for the $W'$ and $Z$ points,  are on the $k_z=0$ plane.
}
\label{fmband}
\end{figure}

The touching point with the quadratic band dispersions in all three directions 
at the $\Gamma$ point can be split by a trigonal distortion,
i.e., by applying a strain or pressure. 
As a result, all bands are separated in the whole BZ.
Varying the $c/a$ ratio of the lattice parameters with a fixed volume 
shifts the touching bands along the $\langle100\rangle$ directions
and coincidentally makes them cross linearly, resulting in a DP.
$c$-axis elongation monotonically shifts the DP along the $Z-\Gamma$ line,
while $c$-axis compression shifts the DP along the $\Gamma-X$ line.
For SOC(001), 
the $Z-\Gamma$ and $\Gamma-X$ lines have $C_{4v}$ and $C_{2v}$ symmetries, respectively.
Thus, inclusion of SOC(001) opens a gap at the point for the compression case, 
whereas the DP survives in the elongation case.\cite{yangbj14,z.gao16}
Such topological transitions in the quadratically touching points 
are very similar to those previously proposed in the fcc Cu$_2$Se.\cite{z.zhu18}

\begin{figure}[tbh]
{\resizebox{8cm}{6.5cm}{\includegraphics{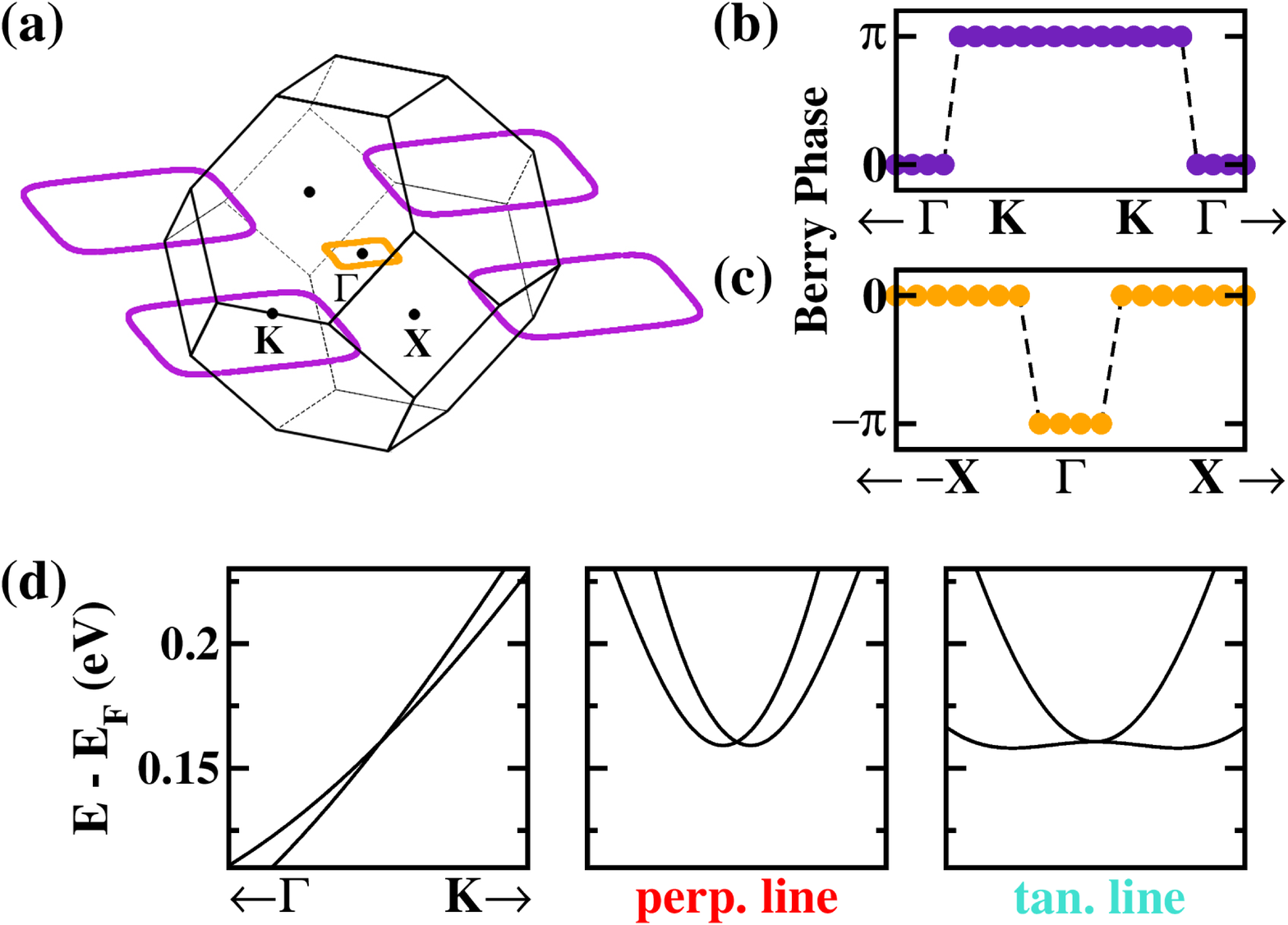}}}
\caption{
(a) Magnetic Weyl nodal loops, appearing on the $M_z$ plane, of the insulating FM \ssoo~ for GGA+SOC(001)+$U$ 
with $U_{eff}=4$ eV.
The BZ-penetrating loops are given only for the filled bands.
(b),(c) Berry phases along the $\Gamma-K$ and the $\Gamma-X$ lines
for closed paths crossing the loops, respectively.
(d) Band dispersions for the $\Gamma$-centered nodal loops
in the perpendicular and tangential directions to the $k_z=0$ plane
in GGA+SOC(001)+$U$ at $U_{eff}=4$ eV.
}
\label{nl}
\end{figure}

\subsection{Ferromagnetic phase}
In the uncorrelated limit, the topological characters of FM \bnoo~are very similar to the results of our model calculations.
In the insulating phase\cite{bnoo,bnoo1},  
\bnoo~shows two MWNLs that emerge only in the unfilled orbitals and
that are similar to those in \ssoo~that will be studied here.
Thus, in this subsection, we concentrate on the FM \ssoo~with a high $T_C$.

Before discussing the observed insulating state,
the correlation was excluded to investigate the sole effects of SOC and broken ${\cal T}$ in this system.
The band structure of GGA+SOC(001) is given in Fig. \ref{fmband}(a).
As previously mentioned,
in addition to the type-I and -II WPs along the $Z-\Gamma$ line,
MWNLs appear just below $E_F$ on the $M_z$ plane.
Other pieces of these MWNLs lie at about 0.85 eV higher in energy, which is close to the strength of $\Delta_{ex}$.


The correlation is now included within the GGA+SOC+$U$ approach to obtain the FM insulating phase.
At the critical value of $U^c_{eff}\approx$3.8 eV, lower by about 2 eV than in the $5d^1$ \bnoo,\cite{bnoo1} 
a relativistic Mott transition occurs in \ssoo.
Figure \ref{fmband}(b) displays the band structure at $U_{eff}=4$ eV, 
where a Mott gap of 0.1 eV is clearly visible.
Remarkably, the combination of mirror symmetry and broken ${\cal T}$
gives rise to several MWNLs robust against SOC on the $M_z$ plane, as shown in Fig. \ref{nl}(a).
Three MWNLs emerge at about --0.5, 0.7, and 1.25 eV, relative to $E_F$, along the $W-K-\Gamma$ line .
Additionally, a MWNL encircling the $\Gamma$ point occurs at $\sim$0.1 eV.
The last MWNL is sensitive to the strength of $U$,
whereas the other MWNLs are nearly insensitive to $U$.
Around $U_{eff}=4$ eV, the $\Gamma$-centered MWNL lies at $\sim$0.1 eV, 
i.e., at the bottom of the conduction band.
Above $U_{eff}=5$ eV, the MWNL lies at $\sim -1$ eV.


To uncover the topological nature of these MWNLs,
the Berry phases were calculated for a closed path crossing these NLs 
on the $k_z=0$ plane.\cite{berryp}
The obtained results are displayed in Figs. \ref{nl}(b) and \ref{nl}(c),
showing jumps from zero to $\pm\pi$, when the MWNLs touch.
This behavior indicates that both NLs are topologically nontrivial.
We also investigated the band dispersions around the nodes 
in the perpendicular and tangential directions to the $k_z=0$ plane.
As shown in Fig. \ref{nl}(d), 
the $\Gamma$-centered loop has an unusual dispersion with tilted crossing bands.
In the perpendicular direction, these two bands touch quadratically  
and have the same sign of effective mass.
These features indicate the type-II NL,\cite{type2nl16,type2nl17,type2nl18,type2nl19} 
where the Lorentz invariance is broken.
On the other hand, the BZ-penetrating loops 
show a usual dispersion (not shown here), indicating the type-I NL.
Thus, this FM phase can be classified as a topological insulator 
with a hybrid (or type-1.5) MWNL, which has not been observed experimentally yet.
These MWNLs would be easily measurable, since they have a small energy dispersion of less than 40 meV.

It is worth noting that the appearance of WPs is sensitive to the choice of the value $U$.
At $U_{eff}=4$ eV, a pair of WPs appears only at 0.7 eV along the $\Gamma-Z$ line [see Fig. \ref{fmband}(b)].

\begin{figure}[tb]
{\resizebox{7cm}{4.5cm}{\includegraphics{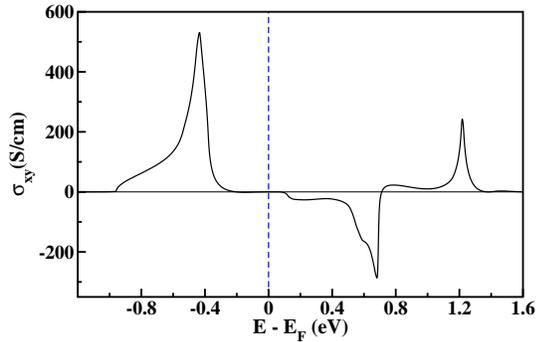}}}
\caption{
Energy dependent anomalous Hall conductivity $\sigma_{xy}$ of \ssoo~
in GGA+SOC(001)+$U$ at $U_{eff}=4$ eV.
The strong anisotropy in the SOC leads to sharp peaks at the energy levels of the MWNLs.
}
\label{ahe}
\end{figure}

\subsection{Anomalous Hall conductivity in the ferromagnetic phase}
As mentioned previously, once SOC is included in the FM insulating phase of \ssoo,
the MWNLs in the mirror-symmetry broken planes become gapped,
whereas those on the $M_z$ plane survive [see Fig. \ref{fmband}(b)].
These SOC-induced small gaps near the energy level of the MWNLs suggest a large Berry curvature.
So, we calculated the AHC $\sigma_{xy}$ from the Berry curvature.\cite{ahc-eq1,ahc-eq2,niu2006}.
The obtained $\sigma_{xy}$ as a function of energy is given in Fig. \ref{ahe}.
As expected, 
there are three sharp peaks at about --0.5, 0.7, and 1.2 eV
where the small SOC gaps appear.
Interestingly, these peaks are isolated at each energy level, making them easily measurable.

In particular, the value of $\sigma_{xy}=530$ S/cm in the occupied band 
is remarkably high, about 30\% larger than the calculated value in the Heusler Fe$_2$Mn$X$ ($X=$ P, As, Sb)\cite{noky2019}
and just half of the experimentally measured value in the FM Co$_3$Sn$_2$S$_2$ 
showing a giant anomalous Hall effect.\cite{liu2018} 
Using the virtual crystal approximation by replacing the atomic number of the Sr ion in the B site, 
our results suggest that this energy level is feasibly achieved by a hole doping of 1.15$\times$10$^{21}$ holes/cm$^3$.
Thus, an AHC measurement is expected to be a useful tool to observe this robust MWNL.  
Alternatively, this topological character could be observed by the anomalous Nernst effect $\alpha_{xy}$,\cite{nernst,hyang2020}
which is proportional to $\frac{d\sigma_{xy}}{dE}$ at $E_F$ in the low-temperature region.\cite{niu2006}

\begin{figure}[tb]
{\resizebox{8cm}{4cm}{\includegraphics{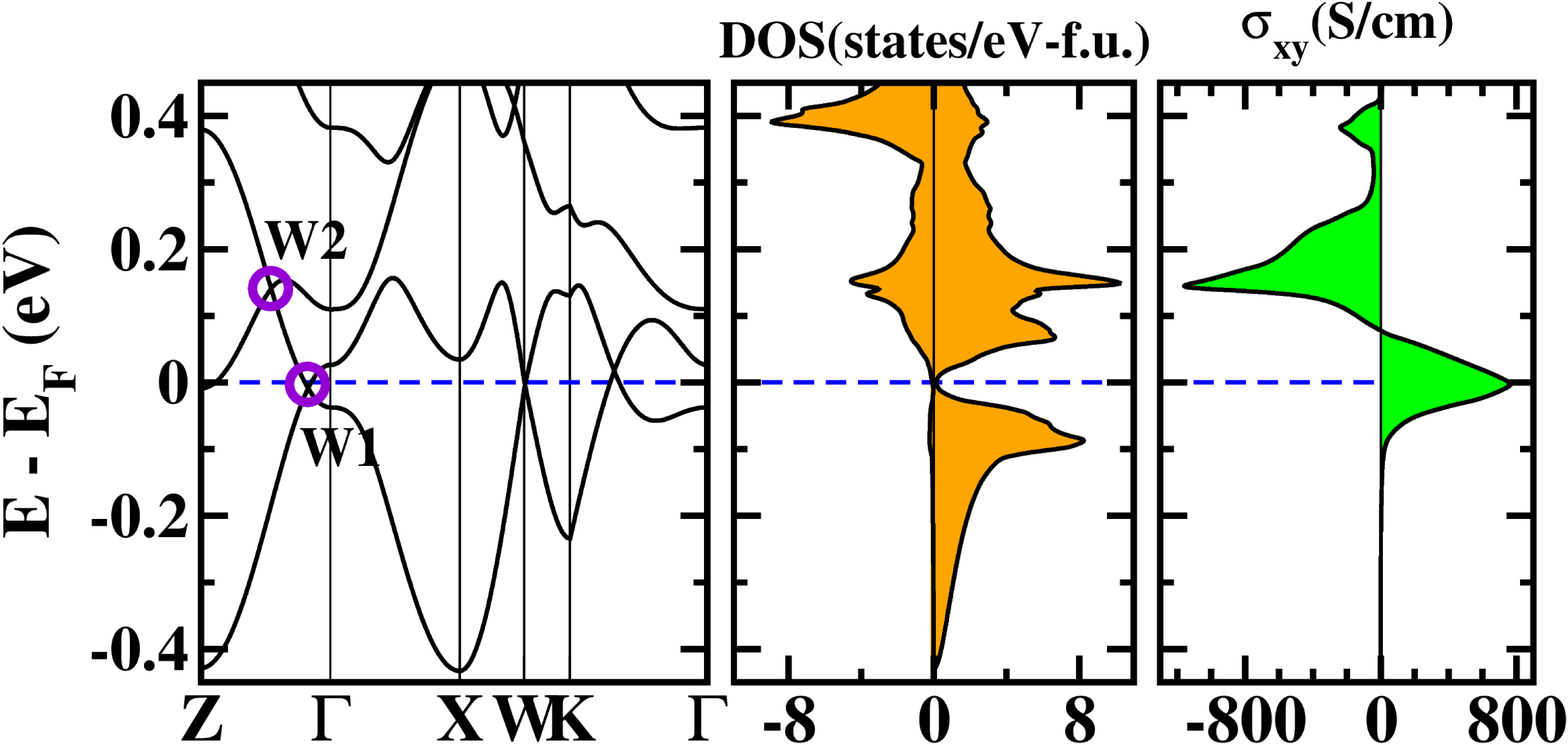}}}
\caption{
Left: GGA+SOC(001) band structure of FM \bzro, enlarged in the range of --0.5 to 0.5 eV.
Two types of WPs, W1 and W2, are denoted by (violet) circles along the $\Gamma-Z$ line.
Other WPs at the $W$ point and along the $\Gamma-K$ line lead to a nodal loop (see text).
Center: The corresponding spin-polarized total DOS,
indicating a half-semimetallic character.
Right: Energy dependent AHC $\sigma_{xy}$, showing two sharp peaks at 0.15 eV and at precisely $E_F$.
}
\label{bs_bzro}
\end{figure}

\section{Ideal Magnetic Weyl Half Semimetals in Ferromagnetic Rhenates}
As can be realized from the hopping parameters given in Table \ref{t1}, 
$E_W\sim E_{\Gamma}$ in \bmro~and \bzro.
This results in very similar topological features of these $5d^1$ rhenates in NM phase 
to those of the NM \bnoo.
Besides, the FM band structures of these rhenates are very similar to each other,
except for more bands crossing $E_F$ in \bmro.
These rhenates commonly show the topological features obtained from our model calculations
of two types of WPs and MWNLs. Thus, we concentrate on the FM phase of \bzro~in this section.

\begin{figure}[tb]
{\resizebox{5cm}{8cm}{\includegraphics{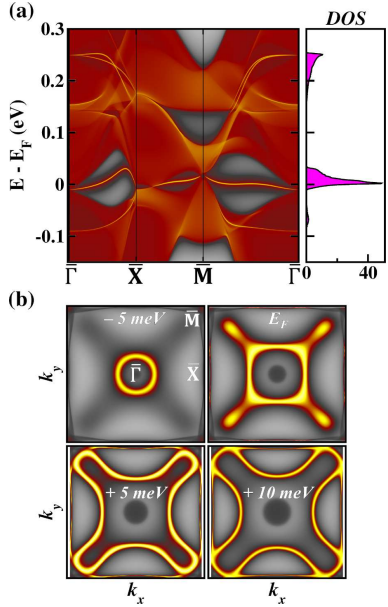}}}
\caption{
(a) (001) surface spectrum and DOS (in arbitrary units) in GGA+SOC(001) of \bzro.
The surface DOS shows a sharp peak at $E_F$ that results from the drumhead state.
(b) Evaluation of surface spectra on isoenergy contours at $E_F$ --5, 0, +5, and +10 (in units of meV). 
The strong yellowish lines denote the surface state.
}
\label{sf_bzro}
\end{figure}

The band structure of FM \bzro~in GGA+SOC(001) 
is displayed in the left panel of Fig. \ref{bs_bzro}. 
The corresponding spin-polarized total DOS is given in the center panel of Fig. \ref{bs_bzro},
showing nearly no contribution of the minority channel in the range --0.5 eV to $E_F$, 
and semimetallic character in the majority channel.
This indicates a half (semi)metal, which has been considered to be a good candidate for spintronics applications.
Returning to the band structure,  
there are, remarkably, only linearly crossing bands near $E_F$ except for the $Z$ point, where a tiny $E_F$ crossing appears.
Nodal points emerge at --4 and 17 meV (relative to $E_F$) along the $\Gamma-K$ line and at the $W$ point,
leading to MWNLs on the $M_z$ plane that are robust even in the presence of SOC(001).
The shape of these MWNLs is the same as the BZ-penetrating one shown in Fig. \ref{nl}(a).
As in the osmates, the calculated Berry phases for the NLs show a $\pi$ jump (not shown here),
indicating a topologically nontrivial character.
The corresponding surface spectrum and DOS are given in Fig. \ref{sf_bzro}(a),
which shows a long drumhead state driven by these MWNLs at $E_F$ [also see the isoenergy contour plot of Fig. \ref{sf_bzro}(b)].
Most strikingly, the MWNLs of \bzro~only exist very close to $E_F$ within the range of a few decades meV,
and are tiny dispersive, leading to a robust drumhead state that induces a sharp peak at $E_F$ in the surface DOS
[see the right panel of Fig. \ref{sf_bzro}(a)].
Figure \ref{sf_bzro}(b) shows the changes of the isoenergy surface spectral contours, as varying the energy 
in the range of $E_F\pm10$ meV.
Above $E_F$, this drumhead state yields a large starfish-like cone crossing almost the whole BZ.

Additionally, along the $\Gamma-Z$ line, two types of WPs, W1 and W2, appear at $E_F$ and 0.15 eV, respectively.
These interesting topological features are suggested to be easily measurable
due to their nearness to $E_F$ with nearly no interruption by other trivial bulk bands
and their robustness under SOC.
Thus, our results indicate that \bzro~is an ideal half semimetal with single-spin NLs and WPs,
which are robust under SOC.

A large AHC $\sigma_{xy}$ can be also expected for the magnetic WP, which leads to a large Berry curvature, 
as observed in the half-metallic FM Co$_3$Sn$_2$S$_2$.\cite{q.wang18}.
The AHC of \bzro~is given in Fig. \ref{bs_bzro}(c).
Two sharp peaks appear precisely at $E_F$ and at about 0.15 eV.
The former peak has the value of 760 S/cm, induced by W1 and some of the tiny SOC gaps on the mirror-symmetry broken planes.
As expected from the large chiral charge,
the multi-WP W2 strikingly leads to a notably large value of --1160 S/cm,
similar to the value of Co$_3$Sn$_2$S$_2$.\cite{q.wang18}. 

These topological characters of these rhenates would be affected by the correlation strength.
Our calculations indicate that $U_{eff}\approx3.3$ (4.0) eV is necessary to obtain a Mott insulating state in \bzro~(\bmro).
(The detailed electronic and magnetic properties will be discussed elsewhere.\cite{song2020})
As observed in several isovalent and isostructural compounds of the double-perovksite A$_2$BReO$_6$,\cite{kato2004}
the resistivities of the rhenates are very sensitive to the A and B ions.
Therefore, even though \bzro~or \bmro~would be insulating (this remains unclear due to lack of available data), 
a (semi)metallic phase can be achieved by a proper choice of the A and B cations,
resulting in the revival of the topological characters.
Further experiments are required to clarify this.

\section{Summary}

Using both the effective three-band model that well describes the isolated $t_{2g}$ manifolds and {\it ab initio} calculations,
we investigated the abundant topological features that emerge due to the interplay among crystalline symmetries, a large SOC,
and a moderate correlation in the cubic FM osmates and rhenates 
: $5d^1$ Ba$_2$NaOsO$_6$, $5d^2$ Sr$_2$SrOsO$_6$, 
and $5d^1$ Ba$_2$$B$ReO$_6$ ($B$= Mg, Zn).

For the osmates, when the time-reversal symmetry (${\cal T}$) is present, 
spinless Dirac nodal loops form the outer nodal chain linked to each other at the $W$ points in the fcc BZ.
When ${\cal T}$ is broken, in the uncorrelated limit 
SOC leads to two types of WPs along the $C_{4v}$ symmetry line
and MWNLs on the mirror planes [say, $M_z$ for the (001) magnetization direction of SOC].
One of the WPs is a multi-WP with the chiral charge of $|\chi|=$2,
whereas the other is a typical type-I WP with $|\chi|=$1.
In addition to these MWNLs, which penetrate the BZ
and are robust to even SOC and correlation strength variation
owing to the combination of mirror symmetry and broken ${\cal T}$,
a type-II MWNL encircling the zone center appears
above the critical on-site Coulomb repulsion $U^c_{eff}$ for the metal-insulator transition.

Presuming (semi)metallic FMs Ba$_2$$B$ReO$_6$ ($B$= Mg, Zn), 
our results suggest that Ba$_2$ZnReO$_6$ is an ideal magnetic Weyl nodal half semimetal 
showing a nearly dispersionless drumhead state at precisely $E_F$ 
and no interference of topologically trivial bulk states even at the large SOC strength.
Thus, this suggests that \bzro~is a promising candidate for novel spintronic applications.

Furthermore, these FM phases lead to the large anomalous Hall conductivities $\sigma_{xy}$,
the maximum value of  $\sim$1160 ($\Omega$cm)$^{-1}$ at 0.15 eV in \bzro.
Finally, our findings suggest that these interesting topological features could commonly emerge 
in many cubic (or slightly distorted) $t_{2g}$ systems 
and their energy levels could be easily tuned by replacement of 5$d$ ion or hole doping. 

\section{Acknowledgments}
We acknowledge Kyo-Hoon Ahn for initial useful collaborations, 
and Youngkuk Kim for fruitful discussions on effects of strain.
This research was supported by  National  Research  Foundation  of  Korea
Grant No. NRF-2019R1A2C1009588.


\begin{thebibliography}{10}

\bibitem{bnoo}
K.-W. Lee and W. E. Pickett, 
Europhys. Lett. {\bf 80}, 37008 (2007)

\bibitem{bnoo1}
S. Gangopadhyay and W. E. Pickett, 
Phys. Rev. B {\bf 91}, 045133 (2015).


\bibitem{wep97} W. E. Pickett, 
Phys. Rev. B {\bf 57}, 10613 (1998).

\bibitem{LP08} K.-W. Lee and W. E. Pickett,
Phys. Rev. B {\bf 77}, 115101 (2008).

\bibitem{pi17} S.-T. Pi, H. Wang, J. Kim, R. Wu, Y.-K. Wang, and C.-K. Lu,
J. Phys. Chem. Lett. {\bf 8}, 332 (2017).


\bibitem{ts1} N. P. Armitage, E. J. Mele, and A. Vishwanath, 
Rev. Mod. Phys. {\bf 90}, 015001 (2018). 

\bibitem{rev-tsm-APX}
S.-Y. Yang, H. Yang, E. Derunova, S. S. P. Parkin, B. Yan, and M. N. Ali,
Adv. Phys. X {\bf 3}, 1414631 (2018).

\bibitem{ts-exp1} 
T. Liang, Q. Gibson, M. N. Ali, M. Liu, R. J. Cava, and N. P. Ong, 
Nat. Mater. {\bf 14}, 280 (2015).

\bibitem{ts-exp2} M. N. Ali, J. Xiong, S. Flynn, J. Tao, Q. D. Gibson, L. M. Schoop, T. Liang, 
N. Haldolaarachchige, M. Hirschberger, N. P. Ong, and R. J. Cava, 
Nature {\bf 514}, 205 (2014).


\bibitem{magWeyl18} L. Ye, M. Kang, J. Liu, F. von Cube, C. R. Wicker, T. Suzuki, C. Jozwiak,
A. Bostwick, E. Rotenberg, D. C. Bell, L. Fu, R. Comin, and J. G. Checkelsky,
Nature {\bf 555}, 638 (2018).

\bibitem{magWeyl19-3} N. Morali, R. Batabyal, P. K. Nag, E. Liu, Q. Xu, Y. Sun, B. Yan, 
C. Felser, N. Avraham, and H. Beidenkopf,
Science {\bf 365}, 1286 (2019).

\bibitem{magWeyl19-2} I. Belopolski, K. Manna, D. S. Sanchez, G. Chang, B. Ernst, J. Yin, 
S. S. Zhang, T. Cochran, N. Shumiya, H. Zheng, B. Singh, G. Bian, D. Multer, M. Litskevich, X. Zhou, 
S.-M. Huang, B. Wang, T.-R. Chang, S.-Y. Xu, A. Bansil, C. Felser, H. Lin, and M. Z. Hasan,
Science {\bf 365}, 1278 (2019).

\bibitem{magWeyl19} T. Suzuki, L. Savary, J.-P. Liu, J. W. Lynn, L. Balents, and J. G. Checkelsky,
Science {\bf 365}, 377 (2019).


\bibitem{WSnoP} S.-Y. Xu, I. Belopolski, N. Alidoust, M. Neupane, G. Bian, 
 C. Zhang, R. Sankar, G. Chang, Z. Yuan, C.-C. Lee, S.-M. Huang,
 H. Zheng, J. Ma, D. S. Sanchez, B. Wang, A. Bansil, F. Chou,
 P. P. Shibayev, H. Lin, S. Jia, and M. Z. Hasan,
 Science {\bf 349}, 613 (2015).


\bibitem{brad16} B. Bradlyn, J. Cano, Z. Wang, M. G. Vergniory, C. Felser, R. J. Cava, and B. A. Bernevig,
 Science {\bf 353}, 11f5037 (2016).

\bibitem{c.fang12} C. Fang, M. J. Gilbert, X. Dai, and B. A. Bernevig,
 Phys. Rev. Lett. {\bf 108}, 266802 (2012).


\bibitem{DW-exp19} D. Takane, Z. Wang, S. Souma, K. Nakayama, T. Nakamura, H. Oinuma, Y. Nakata, H. Iwasawa, C. Cacho, 
 T. Kim, K. Horiba, H. Kumigashira, T. Takahashi, Y. Ando, and T. Sato,
 Phys. Rev. Lett. {\bf 122}, 076402 (2019).

\bibitem{huang16} S.-M. Huang, S.-Y. Xu, I. Belopolski, C.-C. Lee, G. Chang, T.-R. Chang, B. Wang, 
 N. Alidoust, G. Bian, M. Neupane, D. Sanchez, H. Zheng, H.-T. Jeng, A. Bansil, T. Neupert, H. Lin, and M. Z. Hasan,
 Proc. Natl. Acad. Sci. (USA) {\bf 113}, 1180 (2016).

\bibitem{z.zhu18} Z. Zhu, Y. Liu, Z.-M. Yu, S.-S. Wang, Y. X. Zhao, Y. Feng, X.-L. Sheng, and S. A. Yang,
 Phys. Rev. B {\bf 98}, 125104 (2018).

\bibitem{tang17} P. Tang, Q. Zhou, and S.-C. Zhang,
 Phys. Rev. Lett. {\bf 119}, 206402 (2017).


\bibitem{QAHE11} G. Xu, H. Weng, Z. Wang, X. Dai, and Z. Fang,
Phys. Rev. Lett {\bf 107}, 186806 (2011).

\bibitem{jpn02}M. Onoda and N. Nagaosa,
J. Phys. Soc. Jpn. {\bf 71}, 19 (2002).

\bibitem{screw17} S. S. Tsirkin, I. Souza, and D. Vanderbilt,
 Phys. Rev. B {\bf 96}, 045102 (2017).

\bibitem{screw19} W. Luo, X. Wang, and M.-X. Deng,
Solid State Commun. {\bf 300}, 113693 (2019
.

\bibitem{min17} S. Ahn, E. J. Mele, and H. Min,
 Phys. Rev. B {\bf 95}, 161112(R) (2017).
\bibitem{huang17} Z.-M. Huang, J. Zhou, and S.-Q. Shen,
 Phys. Rev. B {\bf 96}, 085201 (2017).

\bibitem{dnl1}
Y. Kim, B. J. Wieder, C. L. Kane, and A. M. Rappe,
Phys. Rev. Lett. {\bf 115}, 036806 (2015).


\bibitem{dnl2}
R. Yu, H. Weng, Z. Fang, X. Dai, and X. Hu,
Phys. Rev. Lett. {\bf 115}, 036807 (2015).


\bibitem{nodalsf1} Q. -F. Liang, J. Zhou, R. Yu, Z. Wang, and H. Weng,
Phys. Rev. B {\bf 93}, 085427 (2016).

\bibitem{nodalsf2} T. Bzdu\v{s}ek and M. Sigrist, 
Phys. Rev. B {\bf 96}, 155105 (2017).

\bibitem{nodalsf3} S. A. Yang, Spin {\bf 6}, 1 (2016).

\bibitem{nodalsf4} C. Zhang, Y. Chen, Y. Xie, S. A. Yang, M. L. Cohen, and S. B. Zhang,   
Nanoscale {\bf 8}, 7232 (2016).

\bibitem{nodalchain1} G. Chang, S.-Y. Xu, X. Zhou, S.-M. Huang, B. Singh, B. Wang, I. Belopolski, J. Yin, 
S. Zhang, A. Bansil, H. Lin, and M. Z. Hasan,
Phys. Rev. Lett. {\bf 119}, 156401 (2017).

\bibitem{nodalchain2} R. Yu, Q. Wu, Z. Fang, and H. Weng,
Phys. Rev. Lett. {\bf 119}, 036401 (2017).

\bibitem{nodalchain3} R. Bi, Z. Yan, L. Lu, and Z. Wang,
Phys. Rev. B {\bf 96}, 201305(R) (2017).

\bibitem{nodalchain4} W. Chen, H.-Z. Lu, and J.-M. Hou,
Phys. Rev. B {\bf 96}, 041102(R) (2017).

\bibitem{twobands}Q. Wu, A. A. Soluyanov, and T. Bzdu\v{s}ek,
Science {\bf 365}, 1273 (2019).

\bibitem{drum} N. B. Kopnin, T. T. Heikkil\"a, and G. E. Volovik, 
Phys. Rev. B {\bf 83}, 220503(R) (2011).

\bibitem{lado19} W. Chen and J. L. Lado,
 Phys. Rev. Lett. {\bf 122}, 016803 (2019).

\bibitem{type2WS} A. A. Soluyanov, D. Gresch, Z. Wang, Q. Wu, M. Troyer, X. Dai, and B. A. Bernevig,
 Nature {\bf 527}, 495 (2015).

\bibitem{type2nl16} F.-Y. Li, X. Luo, X. Dai, Y. Yu, F. Zhang, and G. Chen,
 Phys. Rev. B {\bf 94}, 121105(R) (2016).

\bibitem{type2nl17} S. Li, Z.-M. Yu, Y. Liu, S. Guan, S.-S. Wang, X. Zhang, Y. Yao, and S. A. Yang,
 Phys. Rev. B {\bf 96}, 081106(R) (2017).

\bibitem{type2nl18} J. He, X. Kong, W. Wang, and S.-P. Kou,
 New J. Phys. {\bf 20}, 053019 (2018).

\bibitem{type2nl19} T.-R. Chang, I. Pletikosic, T. Kong, G. Bian, A. Huang, J. Denlinger,
 S. K. Kushwaha, B. Sinkovic, H.-T Jeng, T. Valla, W. Xie, and R. J. Cava,
 Adv. Sci. {\bf 6}, 1800897 (2019).

\bibitem{ts-rot1} 
Z. Wang, Y. Sun, X.-Q. Chen, C. Franchini, G. Xu, H. Weng, X. Dai, and Z. Fang,
Phys. Rev. B {\bf 85}, 195320 (2012).

\bibitem{ts-rot2}
Z. K. Liu, B. Zhou, Y. Zhang, Z. J. Wang, H. M. Weng, D. Prabhakaran, S.-K. Mo, Z. X. Shen, Z. Fang,
 X. Dai, Z. Hussain, and Y. L. Chen,
Science {\bf 343}, 864 (2014).

\bibitem{ts-m1} 
G. Bian, T.-R. Chang, R. Sankar, S.-Y. Xu, H. Zheng, T. Neupert, C.-K. Chiu, S.-M. Huang, G. Chang, 
I. Belopolski, D. S. Sanchez, M. Neupane, N. Alidoust, C. Liu, B. Wang, C.-C. Lee, H.-T. Jeng, 
C. Zhang, Z. Yuan, S. Jia, A. Bansil, F. Chou, H. Lin, and M. Z. Hasan, 
Nat. Commun. {\bf 7}, 10556 (2016).

\bibitem{ts-m2}
G. Bian, T.-R. Chang, H. Zheng, S. Velury, S.-Y. Xu, T. Neupert, C.-K. Chiu, 
S.-M. Huang, D. S. Sanchez, I. Belopolski, N. Alidoust, P.-J. Chen, G. Chang, A. Bansil, 
H.-T. Jeng, H. Lin, and M. Z. Hasan, 
Phys. Rev. B {\bf 93}, 121113(R) (2016).

\bibitem{ts-ns1}
C. Fang, Y. Chen, H.-Y. Kee, and L. Fu, 
Phys. Rev. B {\bf 92}, 081201(R) (2015).

\bibitem{ts-ns2} 
L. M. Schoop, M. N. Ali, C. Stra$\ss$er, A. Topp, A. Varykhalov, D. Marchenko, V. Duppel,
S. S. P. Parkin, B. V. Lotsch, and C. R. Ast, 
Nat. Commun. {\bf 7}, 11696 (2016).


\bibitem{c.fang16} C. Fang, H. Weng, X. Dai, and Z. Fang,
 Chin. Phys. B {\bf 25}, 117106 (2016).

\bibitem{y.kim18} H. Gao, J. W. F. Venderbos, Y. Kim, and A. M. Rappe,
 Annu. Rev. Mater. Res. {\bf 49}, 153 (2019).


\bibitem{magloop1} C. Chen, Z.-M. Yu, S. Li, Z. Chen, X.-L. Sheng, and S. A. Yang,
 Phys. Rev. B {\bf 99}, 075131 (2019).

\bibitem{magloop2} S. Nie, H. Weng, and F. B. Prinz,
 Phys. Rev. B {\bf 99}, 035125 (2019).

\bibitem{magloop3} J. N. Nelson, J. P. Ruf, Y. Lee, C. Zeledon, J. K. Kawasaki, S. Moser,
 C. Jozwiak, E. Rotenberg, A. Bostwick, D. G. Schlom, K. M. Shen, and L. Moreschini,
 Phys. Rev. Materials {\bf 3}, 064205 (2019).

\bibitem{bnoo-exp1}
K. E. Stitzer, M. D. Smith, and H.-C. zur Loye, 
Solid State Sci. {\bf 4}, 311 (2002).

\bibitem{bnoo-exp2} A. S. Erickson, S. Misra, G. J. Miller, R. R. Gupta, Z. Schlesinger, 
W. A. Harrison, J. M. Kim, and I. R. Fisher,
Phys. Rev. Lett. {\bf 99}, 016404 (2007).

\bibitem{ssoo-exp}
Y. K. Wakabayashi, Y. Krockenberger, N. Tsujimoto, T. Boykin, S. Tsuneyuki, Y. Taniyasu, and H. Yamamoto,
Nat. Commun. 10, 535 (2019).

\bibitem{bzro} C. A. Marjerrison, C. M. Thompson, G. Sala, D. D. Maharaj, E. Kermarrec,
 Y. Cai, A. M. Hallas, M. N. Wilson, T. J. S. Munsie, G. E. Granroth, R. Flacau, J. E. Greedan,
 B. D. Gaulin, and G. M. Luke,
 Inorg. Chem. {\bf 55}, 10701 (2016).


\bibitem{nexus19} H.-S. Jin, Y.-J. Song, W. E. Pickett, and K.-W. Lee,
 Phys. Rev. Materials {\bf 3}, 021201(R) (2019).

\bibitem{r.zhang20} R.-W. Zhang, Z. Zhang, C.-C. Liu, and Y. Yao,
 Phys. Rev. Lett. {\bf 124}, 016402 (2020).


\bibitem{gga} 
J. P. Perdew, K. Burke, and M. Ernzerhof
Phys. Rev. Lett. {\bf 77}, 3865 (1996).

\bibitem{wien2k} K. Schwarz and P. Blaha,
Comput. Mater. Sci. {\bf 28}, 259 (2003).


\bibitem{w90}
A. A. Mostofi, J. R. Yates, Y.-S. Lee, I. Souza, D. Vanderbilt, and N. Marzari, 
Comput. Phys. Commun. {\bf 178}, 685 (2008)

\bibitem{w2w}
J. Kune\u{s}, R. Arita, P. Wissgott, A. Toschi, H. Ikeda, and K. Held, 
Comput. Phys. Commun. {\bf 181}, 1888 (2010).

\bibitem{surfgr} 
M. P. L\'{o}pez Sancho, J. M. L\'{o}pez Sancho, and J. Rubio,
J. Phys. F: Met. Phys. {\bf 15}, 851-858 (1985)

\bibitem{wt} 
Q. S. Wu, S. N. Zhang, H.-F. Song, M. Troyer, and A. A. Soluyanov, 
Comput. Phys. Commun. {\bf 224}, 405 (2018).

\bibitem{ahc-eq1} 
D. J. Thouless, M. Kohmoto, M. P. Nightingale, and M. den Nijs,
Phys. Rev. Lett. {\bf 49}, 405 (1982).

\bibitem{ahc-eq2} 
N. Nagaosa, J. Sinova, S. Onoda, A. H. MacDonald, and N. P. Ong,
Rev. Mod. Phys. {\bf 82}, 1539, (2010).

\bibitem{liu2018} E. Liu, Y. Sun, N. Kumar, L. Muechler, A. Sun, L. Jiao, S.-Y. Yang,
D. Liu, A. Liang, Q. Xu, J. Kroder, V. S\"{u}\ss, H. Borrmann, C. Shekhar, Z. Wang, C. Xi, 
W. Wang, W. Schnelle, S. Wirth, Y. Chen, S. T. B. Goennenwien, and C. Felser,
Nat. Phys. {\bf 14}, 1125 (2018).


\bibitem{wcc14} M. Taherinejad, K. F. Garrity, and D. Vanderbilt,
Phys. Rev. B {\bf 89}, 115102 (2014).


\bibitem{chern05} T. Fukui, Y. Hatsugai, and H. Suzuki,
J. Phs. Soc. Jpn. {\bf 74}, 1674 (2005).

\bibitem{pbpd3} K.-H. Ahn, W. E. Pickett, and K.-W. Lee,
  Phys. Rev. B {\bf 98}, 035130 (2018).


\bibitem{teo08}J. C. Y. Teo, L. Fu, and C. L. Kane,
Phys.Rev.B {\bf 78}, 045426 (2008).

\bibitem{tci} L. Fu, 
  Phys. Rev. Lett. {\bf 106}, 106802 (2011).

\bibitem{ttn} M.-C. Jung, K.-W. Lee, and W. E. Pickett,
  Phys. Rev. B {\bf 97}, 121104(R) (2018).


\bibitem{yangbj14} B.-J. Yang and N. Nagaosa,
Nat. Commun. {\bf 5}, 4898 (2014).

\bibitem{z.gao16} Z. Gao, M. Hua, H. Zhang, and X. Zhang,
Phys. Rev. B {\bf 93}, 205109 (2016).


\bibitem{berryp} 
D. Xiao, M.-C. Chang, and Q. Niu,
Rev. Mod. Phys. {\bf 82}, 1959, (2010)


%
\bibitem{niu2006} D. Xiao, Y. Yao, Z. Fang, and Q. Niu,
Phys. Rev. Lett. {\bf 97}, 026603 (2006).


\bibitem{noky2019}
J. Noky, Q. Xu, C. Felser, and Y. Sun,
Phys. Rev. B. {\bf 99}, 165117 (2019).

\bibitem{nernst} 
J. Noky, J. Gooth, C. Felser, and Y. Sun,
Phys. Rev. B. {\bf 98}, 241106(R) (2018).

\bibitem{hyang2020} H. Yang, W. You, J. Wang, J. Huang, C. Xi, X. Xu, C. Cao, M. Tian, Z.-A. Xu, J. Dai, and Y. Li,
Phys. Rev. Materials {\bf 4}, 024202 (2020).



\bibitem{q.wang18} Q. Wang, Y. Xu, R. Lou, Z. Liu, M. Li, Y. Huang, D. Shen, H. Weng, S. Wang, and H. Lei,
Nat. Commun. {\bf 9}, 3681 (2018).


\bibitem{song2020} Y.-J. Song and K.-W. Lee, (unpublished).

\bibitem{kato2004} H. Kato, T. Okuda, Y. Okimoto, Y. Tomioka, K. Oikawa, T. Kamiyama, and Y. Tokura,
Phys. Rev. B {\bf 69}, 184412 (2004).






\end{thebibliography}
\end{document}